\documentclass[a4paper,fleqn,usenatbib]{mnras}

\usepackage{newtxtext,newtxmath}
\usepackage{nicefrac}
\usepackage[T1]{fontenc}
\usepackage{ae,aecompl}

\usepackage{graphicx}	
\usepackage{amsmath}	
\def\rosat{\emph{ROSAT}}
\def\xmm{\emph{XMM--Newton}}
\def\planck{\emph{Planck}}
\def\kmsMpc{km~s$^{-1}$~Mpc$^{-1}$}

\title[ROSAT soft excess in Coma]{A WHIM origin for the soft excess emission in the Coma cluster}

\author[M. Bonamente et al.]{
  Massimiliano Bonamente,$^{1}$\thanks{E-mail: bonamem@uah.edu (MB)}
  Mohammad Mirakhor,$^{1}$
Richard Lieu,$^{1}$
and Stephen Walker$^{1}$
\\
$^{1}$ Department of Physics and Astronomy, University of Alabama in Huntsville,
Huntsville, AL 35899
}

\date{Accepted XXX. Received YYY; in original form ZZZ}

\pubyear{2022}

\begin{document}
\label{firstpage}
\pagerange{\pageref{firstpage}--\pageref{lastpage}}
\maketitle

\begin{abstract}
This paper provides a new analysis of \rosat\ observations of the Coma cluster, to determine
the amount of soft X--ray radiation in excess of the
contribution from the hot intra--cluster medium. The re--analysis is made possible by a high--resolution
  study of the hot intra--cluster medium with the \xmm\ and \planck\ telescopes
out to the cluster's virial radius. The analysis confirms the original findings of a strong
excess of soft X--ray radiation, which is likely to be of thermal origin.
  We find quantitative agreement between the detected soft excess and the 
  physical characteristics of warm--hot intergalactic medium (WHIM)
filaments seen in hydrodynamical simulations. We conclude that the most plausible explanation for the soft excess is 
  the presence of $\sim 10$~Mpc--long filaments at $\log T(K)\simeq 6$, with a baryon overdensity of
  $\sim 300$, converging towards the Coma cluster. This interpretation therefore provides support for the 
  identification of the missing low--redshift baryons with WHIM filaments, as predicted by numerical simulations.
\end{abstract}

\begin{keywords}
  galaxies: clusters: individual: Coma cluster -- galaxies: clusters: intracluster medium -- cosmology: large--scale structure of the Universe
\end{keywords}



\section{Introduction}

Galaxy clusters are strong emitters of X--ray radiation, which originates from 
a diffuse intergalactic medium at temperatures of $\log T(K) \simeq 7-8$. 
The Coma cluster, also known as Abell 1656, is one of the best-studied clusters 
of galaxies in all energy bands from radio to X-rays. Its proximity 
\citep[$z = 0.0231$;][]{struble1999} and mass \citep[$M_{200} \sim 8.5 \times 10^{14}$ M$_\odot$;][]{mirakhor2020} make it an ideal target  for many studies. X-ray observations have shown many spatial 
features of its intracluster medium, with a number of  infalling substructures such as NGC~4839 
\citep[e.g.][]{Neumann2001,Neumann2003},  signs of surface brightness fluctuations 
\citep[e.g.][]{Churazov2012}, and variations of temperature \citep[e.g.][]{Watanabe1999,simionescu2013}. 
The Coma cluster also features a giant radio halo that extends over scales of about 1 Mpc, 
tracing the non-thermal inverse Compton emission from relativistic electrons and large-scale 
magnetic fields. Furthermore, the \textit{Planck} observation of the Coma cluster 
\citep{PlanckCollaboration2013} revealed 
at least two shock fronts in two separate locations around 40 arcmin to the south-east and west of the Coma center, 
which corresponds roughly to the outer edge of the giant radio halo.
At the Coma redshift, one arcmin corresponds to a distance of 28~kpc, for a Hubble constant of $H_0=70$~\kmsMpc\ and  a
standard $\Lambda$CDM cosmology with $\Omega_m=0.3$ and $\Omega_{\Lambda}=0.7$.

The Coma cluster was one of two clusters where an excess of soft X--ray radiation, above the contribution
from the hot intra--cluster medium (ICM), was originally detected by the \emph{EUVE} mission \citep{lieu1996a,lieu1996b,bowyer1996}.
This excess of radiation, usually referred to as the \emph{cluster soft excess}, was subsequently
confirmed by \rosat\ for the Coma cluster \citep{bonamente2003,bonamente2009} and for several other
nearby clusters \citep{bonamente2002,bonamente2001}, and by several other X--ray instruments including
\xmm\ \citep[e.g.][]{kaastra2003,nevalainen2003,nevalainen2007} and
\emph{BeppoSAX} \citep{bonamente2001b}. Some of the early \emph{EUVE} 
measurements were subject to re--analyses that revised downward some of the soft excess 
fluxes \citep[e.g.][]{bowyer1999},
but there remain a preponderance of evidence that the cluster soft excess is a genuine astrophysical phenomenon.

The origin of this excess of soft X--ray radiation 
has not been determined conclusively, given the limited spectral resolution of the available X--ray spectrometers.
Possible explanations include a non--thermal origin as inverse--Compton scattering 
of cluster cosmic rays with the thermal medium \citep[e.g.][]{lieu1999b,bonamente2005},
or a thermal origin as a sub--virial gas at million--Kelvin temperatures \citep[e.g.][]{lieu1996a,lieu1996b,bonamente2003}. 
The numerical simulations of \cite{cheng2005} provided evidence that cluster may contain
a halo of \emph{warm} sub--virial baryons that can generate the soft excess emission, with a preferential
distribution towards large radii, consistent with the distribution of the excess in many clusters,
including Coma. The thermal origin of the soft excess is also consistent with the presence
of sub--virial gas at the outskirts of clusters, where filaments of warm--hot intergalactic medium (WHIM)
are expected to stretch between groups and clusters of galaxies. Such filaments 
are seen in a range of hydrodynamical
simulations, from the early \cite{cen1999,dave2001} results, to the more recent \texttt{EAGLE}
\citep[e.g.][]{schaye2015,tuominen2021} and  \texttt{IllustrisTNG} simulations \citep{martizzi2019,gallarraga2021}.

The two major X--ray telescopes of the past decades, \xmm\ and \emph{Chandra}, have had a limited
ability to further investigate and clarify the origin of the soft excess emission. This is given 
by the fact that the soft X--ray band of interest
of their spatially--resolved spectrometers 
(photon energies below 1~keV) have low spectral resolution, high background levels and known calibration issues
\citep[see, e.g.][]{bonamente2011}, whereas \rosat\ had a stable and well--calibrated
$\nicefrac{1}{4}$~keV band \citep[e.g.][]{snowden1994} with very low detector background, ideal for low surface--brightness soft X--ray emission. 
The \emph{eRosita} instrument \citep{meidinger2020} onoboard the \emph{SRG} mission has the potential to
investigate  this phenomenon, due to its wide field of view and
soft X--ray response. It is therefore of interest to revisit the soft excess emission in one of the clusters
with strongest signal, ahead of the analysis of \emph{eRosita} observations of Coma {in search of the soft excess
radiation}, and possibly with future 
high spectral resolution instruments such as \textit{XRISM} \citep{xrism2020}. {An initial analysis of the 
\emph{eRosita} Coma data was presented by \cite{churazov2021}, with emphasis on the cluster's spatial
morphology that did not 
address the soft X--ray emission. Another \emph{eRosita} observation of a WHIM filament associated with the Abell~3391/3395 system by \cite{reiprich2021}
shows that its soft band can in fact detect the type of faint emission discussed in this paper.}

This paper is structured as follows. Section~\ref{sec:observations} describes the \rosat, \xmm\ and \planck\ observations
of Coma, Sect.~\ref{sec:rescaling}  presents our method of re--analysis of the \rosat\ data with the newer
\xmm/\planck\ temperatures, Sect.~\ref{sec:results} provides the results and interpretation
of the soft excess in the Coma cluster, and Sect.~\ref{sec:conclusions}
presents our discussion and  conclusions.

\section{Observations of the Coma cluster}
\label{sec:observations}
{This section presents the observations of the Coma cluster used to determine the soft excess fluxes.
The \rosat\ data of \cite{bonamente2003} are  used in conjunction with a more recent
determination of the cluster temperatures by \cite{mirakhor2020}.}

{\subsection{The \rosat\ observations of \protect\cite{bonamente2003}}}
In \cite{bonamente2003} we analyzed several pointed \rosat\ {PSPC} observations out to a radial distance of $\sim$~1.5~degrees,
from which  we reported the presence of strong
{soft excess} emission above the contribution from the hot ICM. 
One of the key limitations in the determination of the soft excess fluxes
was the unavailability of high--resolution temperature maps for the hot ICM, given the
narrow \rosat\ bandpass ($\sim 0.2-2$~keV). In our earlier \rosat\ analysis,
we obtained estimates of the hot ICM from a single--temperature fit to an
optically--thin plasma emission model (\texttt{MEKAL}) in the 1-2~keV band.
These estimates were then extrapolated to the neighboring soft X--ray bands R2--R4 \citep[as 
defined in][ indicated as R24, $\sim 0.2-1$~keV]{snowden1998}
to determine the contribution from the hot ICM in that band.
The \rosat\ temperature profile from \cite{bonamente2003} is reproduced 
in Table~\ref{tab:tprofComp}. The azimuthal coverage of the \rosat\ data is complete out to 55~arcmin,
and it is partial beyond it. A full azimuthal coverage of Coma was also available via the \rosat\ All--Sky Survey data,
which was analyzed in \cite{bonamente2009} to confirm the presence of soft excess. 
Those short--exposure data however were only used for photometric purposes, and they are not considered in this paper.

{\subsection{Systematic errors associated with the \rosat\ analysis}}
{
  Prior to using the \rosat\ soft X--ray fluxes with the \xmm/\planck\ temperatures of \cite{mirakhor2020}, it is
useful to address certain sources of systematic errors associated with the \cite{bonamente2003} analysis of the \rosat\ data.

The soft X--ray fluxes in the R24 band, and especially in the softest $\nicefrac{1}{4}$~keV band (i.e., R2), are
significatly modified by  Galactic absorption. The column density of hydrogen towards Coma used in the
\cite{bonamente2003} paper was measured by 
\cite{dickey1990} and \cite{hartmann1997}, the latter part of the LAB survey \citep{kalberla2005}.
These measurements were
complemented by an analysis of the IRAS data that confirmed a distribution with $N_H=0.9-1.1 \times 10^{20}$~cm$^{-2}$
within 5 degrees of the cluster's center
(see Figure~3 therein). A more recent measurent by the HI4PI survey \citep{HI4PI2016} confirms a smooth
distribution of $N_H$ towards Coma, with an average value of 
$N_H=0.9\pm 0.1 \times 10^{20}$~cm$^{-2}$ towards the central degree of Coma, slightly rising to an average 
value of $N_H=1.1\pm 0.1 \times 10^{20}$~cm$^{-2}$ within a radius of 3 degrees, consistent with the previous measurements
within the statistical uncertainities. The soft excess fluxes measured by \cite{bonamente2003} already include 
the systematic error corresponding to a possible $\pm 1 \times 10^{19}$~cm$^{-2}$ uncertainty in the determination of
the Galactic column density. Given that all available measurements towards the Coma cluster are statistically
consitent with one another, no further
source of systematic error is assessed in the re--analysis of the soft excess fluxes provided below in Sect.~\ref{sec:rescaling}.
Moreover, the cross--sections for the Galactic photoelectric absorption are those of \cite{morrison1983} 
(\texttt{wabs}), which are
in agreement with those of \cite{yan1998} and of the higher--resolution calculations \cite{wilms2000} (\texttt{tbabs}).
An extensive discussed of the photoelectric cross--sections for soft X--ray fluxes is
provided in \cite{bonamente2002} and \cite{bonamente2001c}.~\footnote{The \cite{balucinska1992}  
(\texttt{phabs})  cross--sections  of He are believed to be less accurate than those of either
\cite{morrison1983} or \cite{wilms2000}, with a slightly lower value that would in fact lead to even larger soft
excess fluxes.}

The calibration and stability of the \rosat\ PSPC detector are discussed in \cite{snowden1994} and \cite{snowden2001}, 
including the calibration of the radial dependence of the detector efficiency with off--set angle. A study of the
soft emission from white dwarfs and neutron stars by \cite{beuermann2006} shows that the \rosat\ PSPC soft X--ray fluxes are
well calibrated within $\leq 10$~\%, as also indicated by \cite{snowden1995}. Moreover, the
\cite{bonamente2003} analysis excluded the portions of the detector shadowed by the `wagon wheel' structure used to
house the filters, in correspondence of which the response of the PSPC is significantly reduced.
The \cite{beuermann2006} study also shows the overall good agreement between the \emph{Chandra} HRC and \rosat\ PSPC
instruments, and \cite{snowden2002} finds a $\leq 10$~\% error in the cross--calibration among the \emph{Chandra}, \emph{XMM}
and \rosat\ imaging instruments. These results point to
a stable and well--calibrated response for the PSPC, also when compared to other X--ray instruments.
More importantly, 
an absolute calibration of the \rosat\ and \xmm\ fluxes is in fact 
not crucial for our analysis, 
since only the \xmm/\planck\ temperatures
(and not the fluxes or spectral normalizations) are applied to the \rosat\ data. 

Finally, the \rosat\ PSPC detector was designed with a particle anti--coincidence shield that 
provides a $\geq 99$~\% rejection of particle background \citep{snowden1992,plucinsky1993},
thus making the particle background essentially unimportant for the majority of observations.
The large field--of--view of ROSAT (approximately 1 degree in radius) makes it also possible to
measure \emph{in situ} background at the same time as the cluster observations, in order to
control the time--variable components of the photon background, such as those due by 
charge--exchange. Additional details on the reduction and background subtraction
are provided in \cite{bonamente2002}.
}

{\subsection{The \xmm/\planck\ observations of \protect\cite{mirakhor2020}}}
Combining a large \textit{XMM-Newton} mosaic with the \textit{Planck} 
Sunyaev--Zeldovich effect observations of the Coma cluster, 
\citet{mirakhor2020} studied the thermodynamic properties of the Coma cluster in an 
azimuthally averaged profile and in 36 angular sectors out to the virial radius, 
with nearly full azimuthal coverage. They found that the temperature profiles in azimuthal 
sectors exhibit similar radial trends, with the temperature dropping from about 8.0~keV 
at the Coma core to about 3.0~keV in the outskirts. 
Beyond $r_{500}$, the temperature profiles, on average, tend to be flatten and do not drop with radius. 
Within a radius of 30 arcmin, there is a gradient in the gas temperature from the 
hotter region in the north-west to the colder region in the south-east, 
agreeing with previous measurements \citep[e.g.][]{Watanabe1999,Neumann2003}. 
Along the less disturbed directions, \citet{mirakhor2020} found that the entropy 
measurements follow the power-law entropy profile predicted by non-radiative simulations 
for purely gravitational hierarchical structure formation. However,  an entropy deficit is 
found in the outskirts along the south-west direction, where Coma connects to Abell~1367 
through the cosmic web filament. This entropy deficit extends from $0.5r_{200}$ out to the 
virial radius, consistent with what is expected from simulations of a 
filamentary gas streams that can penetrate  deep into the cluster, bringing 
low-entropy gas into the cluster core.

\begin{figure}
  \includegraphics[width=3.5in]{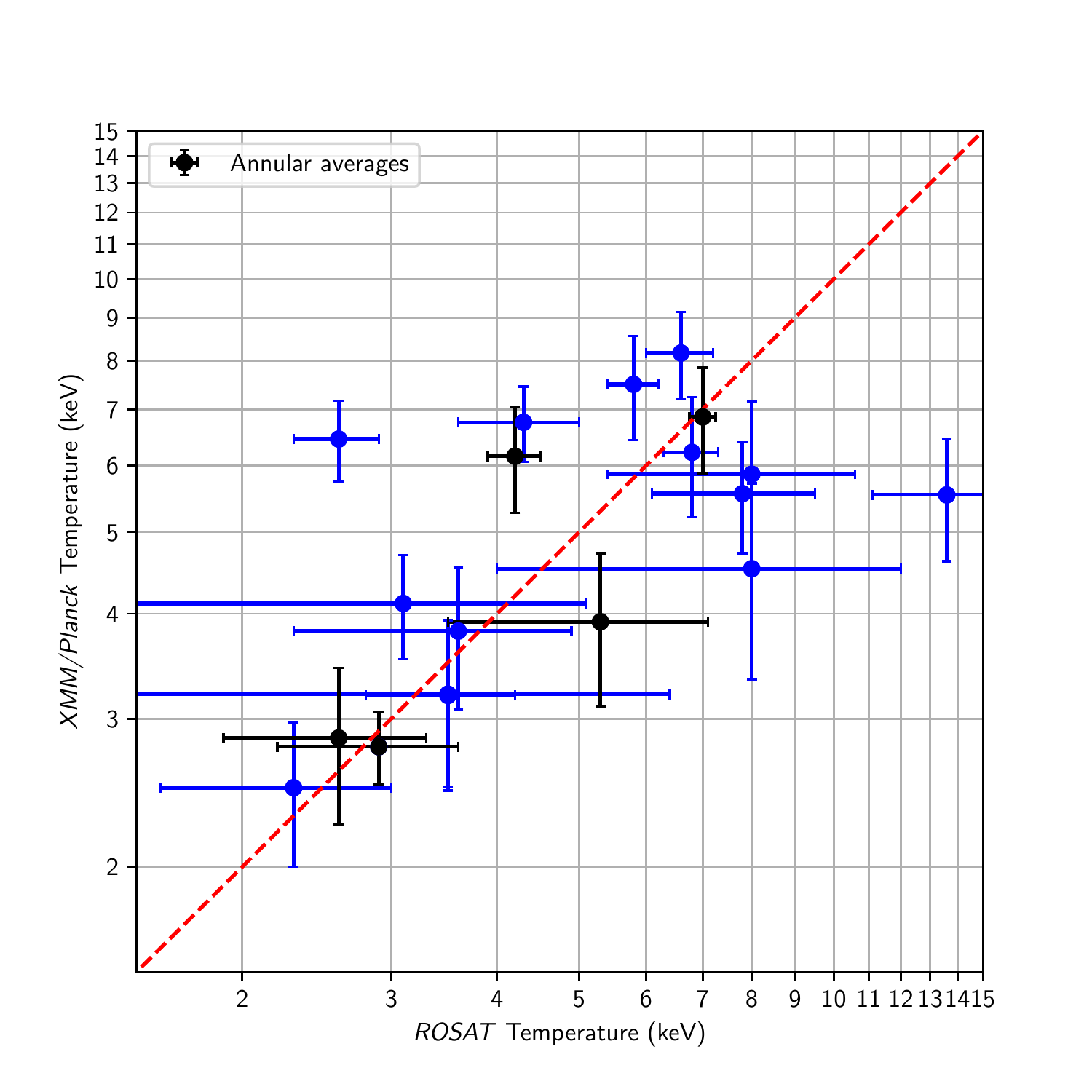}
  \caption{Comparison of \rosat\ and \xmm/\planck\ temperatures in the Coma cluster.
  The blue datapoints correspond to the quadrants in Table~\ref{tab:tprofComp}, in black are the last
  5 datapoints with the azymuthal averages.}
  \label{fig:tprof}
\end{figure}

These newer \xmm/\planck\ measurements make it possible to 
re--analyze the \rosat\ soft X--ray data in light of these more accurate ICM temperatures.
Since the goal of this study is to determine the soft excess flux as measured
by \rosat, we use the averaged \cite{mirakhor2020} temperature measurements as the temperature
value that applies to the \rosat\ radial bins shown in Table~\ref{tab:tprofComp}.
Such averaging procedure leads to the values reported in {the rightmost column of} Table~\ref{tab:tprofComp}.
A comparison of the temperature measurements is {also} reported in Figure~\ref{fig:tprof}.


\begin{table}
  \centering
  \caption{Comparison of \xmm/\planck\ and \rosat\ temperatures. {\rosat\ temperatures are reproduced from \protect\cite{bonamente2003} }. }
  \label{tab:tprofComp}
  \begin{tabular}{llcc}
    \hline
    \hline
    Quadrant & Radius & PSPC $kT$ & EPIC $kT$ \\
	     &  (arcmin)          & (keV) & (keV) \\
     \hline
	 NW & $10.0\pm10.0$ &   $5.80\pm0.40$ & $7.50\pm1.06$\\
 NW & $30.0\pm10.0$ &   $2.60\pm0.30$ & $6.45\pm0.71$\\
 NW & $47.5\pm7.5$ &   $3.10\pm2.00$ & $4.11\pm0.58$\\
 NW & $62.5\pm7.5$ &   $3.50\pm2.90$ & $3.21\pm0.72$\\
 NE & $10.0\pm10.0$ &   $6.60\pm0.60$ & $8.17\pm0.97$\\
 NE & $30.0\pm10.0$ &   $4.30\pm0.70$ & $6.76\pm0.69$\\
 NE & $47.5\pm7.5$ &   $8.00\pm4.00$ & $4.52\pm1.19$\\
 SE & $10.0\pm10.0$ &   $6.80\pm0.50$ & $6.22\pm1.02$\\
 SE & $30.0\pm10.0$ &   $8.00\pm2.60$ & $5.86\pm1.29$\\
 SE & $47.5\pm7.5$ &   $3.60\pm1.30$ & $3.81\pm0.73$\\
 SW & $10.0\pm10.0$ &   $13.60\pm2.50$ & $5.54\pm0.92$\\
 SW & $30.0\pm10.0$ &   $7.80\pm1.70$ & $5.56\pm0.84$\\
 SW & $47.5\pm7.5$ &   $3.50\pm0.70$ & $3.20\pm0.73$\\
 SW & $62.5\pm7.5$ &   $2.30\pm0.70$ & $2.48\pm0.48$\\
 SW & $80.0\pm10.0$ &   $2.90\pm0.70$ & $2.78\pm0.28$\\
  \hline
  & $10.0\pm10.0$ & $7.00\pm0.25$ & $6.86\pm0.99$\\
  & $30.0\pm10.0$ & $4.20\pm0.30$ & $6.16\pm0.88$\\
  & $47.5\pm7.5$ & $5.30\pm1.80$ & $3.91\pm0.81$\\
  & $62.5\pm7.5$ & $2.60\pm0.70$ & $2.85\pm0.60$\\
  & $80.0\pm10.0$ & $2.90\pm0.70$ & $2.78\pm0.28$\\

    \hline
    \hline
  \end{tabular}
\end{table}

\section{Re--analysis of \rosat\ soft excess fluxes with \xmm/\planck\ temperatures}
\def\new{\text{new}}
\label{sec:rescaling}

The soft excess flux is defined as the difference between the measured
\rosat\ {PSPC} R2--R4 band flux ($F_{24}$), covering approximately the 0.2-1~keV energy range
\citep{snowden1994,snowden1998}, and the flux in the same band predicted
from the hot ICM as measured from the neighboring R5--R7 band (1-2~keV) and indicated as $P_{24}$, such that
\[ E_{24}=F_{24}-{P}_{24}.\]
The total measured \rosat\ $F_{24}$ flux is independent of the temperature assumed for the hot ICM,
while the $P_{24}$ prediction is sensitive to the spectral modelling, and primarily the temperature of the hot ICM.
To determine the new soft excess fluxes using the \xmm/\planck\ temperatures,
it is therefore necessary to re--scale
the predicted R24 band flux ${P}_{24}$ to account for the new \xmm/\planck\ temperatures, as described in the following.

\subsection{Determination of  soft--band predictions using \xmm/\planck\ temperatures}
\label{subsec:rescaling}
A few steps are required to rescale the soft excess fluxes to account for the
more accurate \xmm/\planck\ temperatures.
First, the emission integral $I$ of a spectrum is proportional to the square of the
plasma density $n_e^2$ as
\begin{equation} I =\int n_e^2 dV \propto \dfrac{K}{\Lambda(kT,A)},
\label{eq:I}
\end{equation}
and it is proportional to the normalization constant $K$
of the thermal model and the
emissivity or cooling function $\Lambda(kT,A)$ of the plasma. {For this analysis, $I$ indicates the hot ICM emission integral
as determined from the} 
the 1-2~keV \rosat\ spectral {fits using the} \texttt{MEKAL} {code},
\citep{mewe1985,mewe1986,kaastra1992}. {These fits were also} used to determine the prediction $P_{24}$ {
by the same model in the lower--energy R24 band}.

\begin{figure}
  \centering
  \includegraphics[width=3.5in]{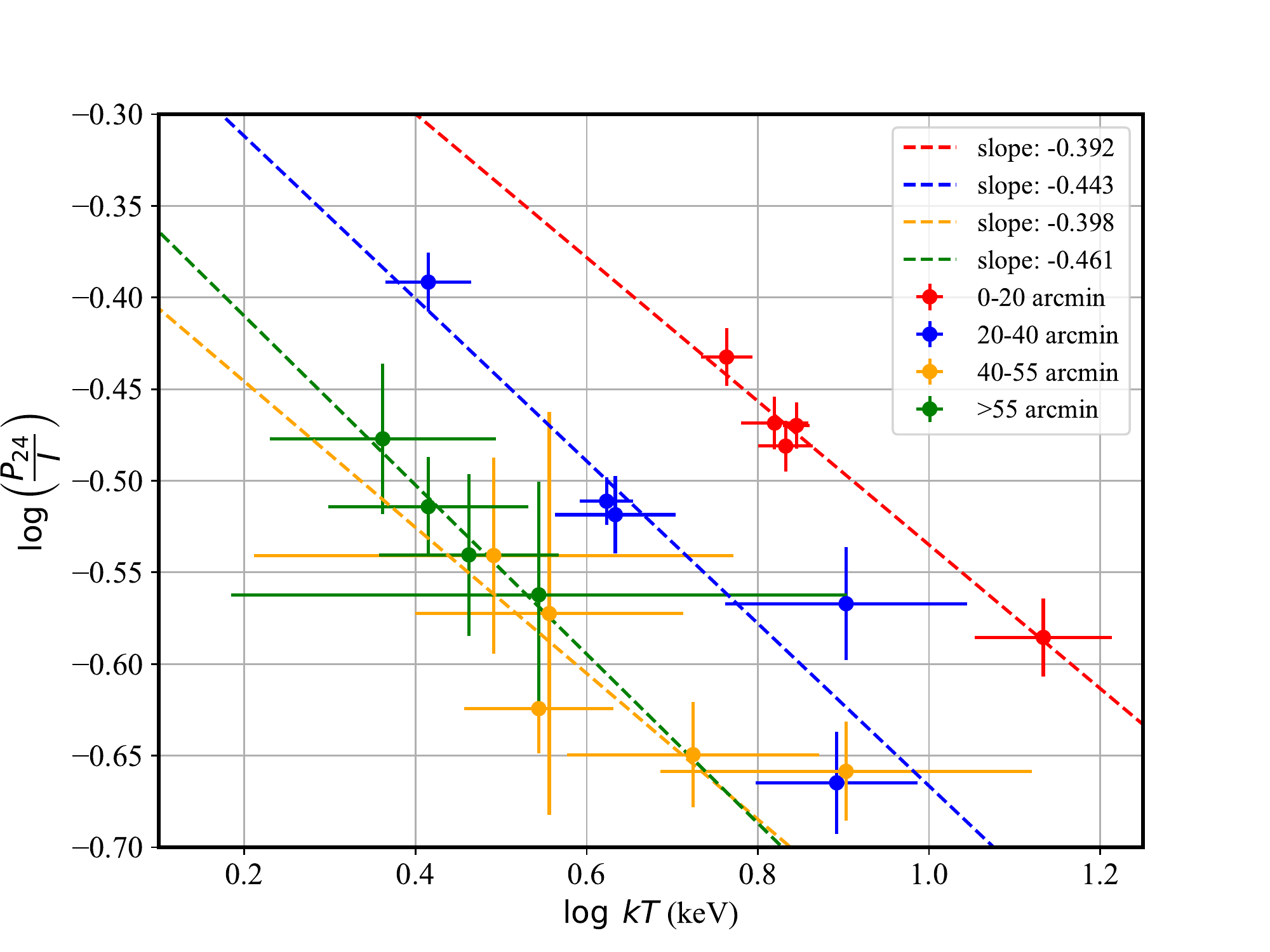}
  \caption{Temperature dependence of the ratio of R24 flux prediction 
  and emission integral, for the \rosat\ data. This scaling relation
  is radius--specific, and it
  accounts for effects of the detector efficiency.}
  \label{fig:predVskT}
\end{figure}

In turn, the measured flux in the 1-2~keV band
is also proportional
to the emission integral, given that detector characteristics such as extraction region size, 
calibration, and efficiency remain constant for a given spectrum, as the temperature of the hot ICM
is varied. We therefore found a simple {empirical} relationship between the ratio of the predicted R24 band flux, and 
the emission integral and the temperature, as
\begin{equation}
  \log \dfrac{P_{24}}{I} = a + b \log kT.
  \label{eq:P24IkT}
\end{equation}
This relationship applies with different best--fit values $a$ and $b$ for different regions,
since the efficiency of the \rosat\ detector is a function of the radial distance from the 
on--axis position. The high accuracy of these relationships can be evinced from Figure~\ref{fig:predVskT},
{ where each point corresponds to the results (i.e., emission integral, prediction in R24 band, and temperature)
of a spectral fit to the \cite{bonamente2003} PSPC spectra.}

The relationship \eqref{eq:P24IkT} can be equivalently rewritten as a function of ratios
between quantities,
\begin{equation}
  \log \dfrac{P_{24,\new}}{P_{24}} = \log \dfrac{I_{\new}}{I} + b \log \dfrac{kT_{\new}}{kT},
  \label{eq:P24IkT2}
\end{equation}
where `\new' refers to the values with the modified temperature. Notice that the parameter $a$ of 
\eqref{eq:P24IkT} is not necessary for \eqref{eq:P24IkT2}, since the
latter uses logarithmic ratios.
Equation \ref{eq:P24IkT2}
is used  to find the new $P_{24}$ prediction \emph{without} the need to fit
again the \rosat\ spectra, but instead using directly the results provided in \cite{bonamente2003}.
Moreover, this equation gives us an opportunity to study analytically the impact of temperature changes
in the hot ICM on the soft excess fluxes, and it provides the  basis for error propagation.
{The emission integral $I$ in the 1-2~keV band is assumed to be due primarily to the hot ICM.
This assumption is justified by the exponential cutoff of the bremsstrahlung emission from 
the warm phase at 
photon energies $\epsilon > kT_w$, whereas the \cite{bonamente2003} analysis indicates typical
warm gas temperatures well below 1~keV. The spectral resolution of the PSPC detector has a FWHM of $\sim 0.4$~keV at
1~keV photon energies (or a standard deviation of ~$\sim 0.17$~keV). This implies that, e.g., even photons at 0.75~keV from 
the high--energy tail of a warm plasma will only have small ($\leq 5$~\%) 
probability of being redistributed into the 1-2~keV band.
}

 In \eqref{eq:P24IkT2} the emission integral (and similarly density and mass) 
will change with the temperature because of the temperature--dependence of the
emissivity  of the plasma. This dependence is approximately described by 
the consideration that the surface brightness is
\[ S_x \propto \int \Lambda(kT,A)\, n_e^2 dV  \propto kT^{-1/2} e^{-E/kT} n_e^2 = \text{const},\]
accounting for the fact that the bulk of the main--band X--ray emission is from thermal bremsstrahlung,
and that the measured surface brightness $S_X$ or flux is the invariant. 
With $E\simeq0.2-1$~keV the energy of the soft X--ray photons in the narrow band of concern, 
and with $kT \geq 2$~keV the hot plasma temperature of Coma, the
dependence of the emission integral on the temperature is approximately
\[ n_e^2 \propto kT^{1/2}.\]
As a result, when a new temperature is introduced in the analysis of the \rosat\ data,
the emission integral must be accordingly modified as
\begin{equation}
  \dfrac{I_{\new}}{I} = \left(\dfrac{n_{e,\new}}{n_e}\right)^2 = \left(\dfrac{kT_{\new}}{kT}\right)^{1/2}.
  \label{eq:IkT}
\end{equation}
This means that if \rosat\ temperatures are revised upwards, so are the corresponding emission
integrals.
 
Finally, use of \eqref{eq:P24IkT2} and \eqref{eq:IkT} lead to the sought--after 
 change in the predicted fluxes in the R24 band,
\begin{equation}
  \dfrac{P_{24,\new}}{P_{24}} = \left(\dfrac{kT_{\new}}{kT}\right)^{\nicefrac{1}{2}+b}
  \label{eqc}
\end{equation}
which, for small changes in $kT$, \eqref{eqc} is approximately
 \begin{equation*}
   \dfrac{\Delta P_{24}}{P_{24}} \simeq \left(b +\dfrac{1}{2}\right)  \dfrac{\Delta kT}{kT} \text{ (small changes)},
 \end{equation*}
 Notice that $b\simeq-0.4$ according to the results shown in Figure~\ref{fig:predVskT}, 
 and therefore the change in the predicted flux is only a very mild function of the temperature
 assumed  for the hot ICM. This finding alone is sufficient to show that even moderately
 large changes in the assumed temperatures, $\Delta kT/kT \leq 1$, lead to 
 changes in the predicted R24 band fluxes, and therefore soft excess fluxes, at the level of $\leq 10\%$.
 This is due to two opposing effects, e.g., an increase in the emission integral as $kT$ is modified upwards
 according to \eqref{eq:IkT}, 
 followed by a reduction in the predicted flux (at constant $I$) as a function of temperature, according to
 \eqref{eq:P24IkT2}.

\subsection{The soft excess flux in \rosat\ with \xmm/\planck\ temperatures and error analysis}

\begin{figure*}
  \includegraphics[width=5.5in]{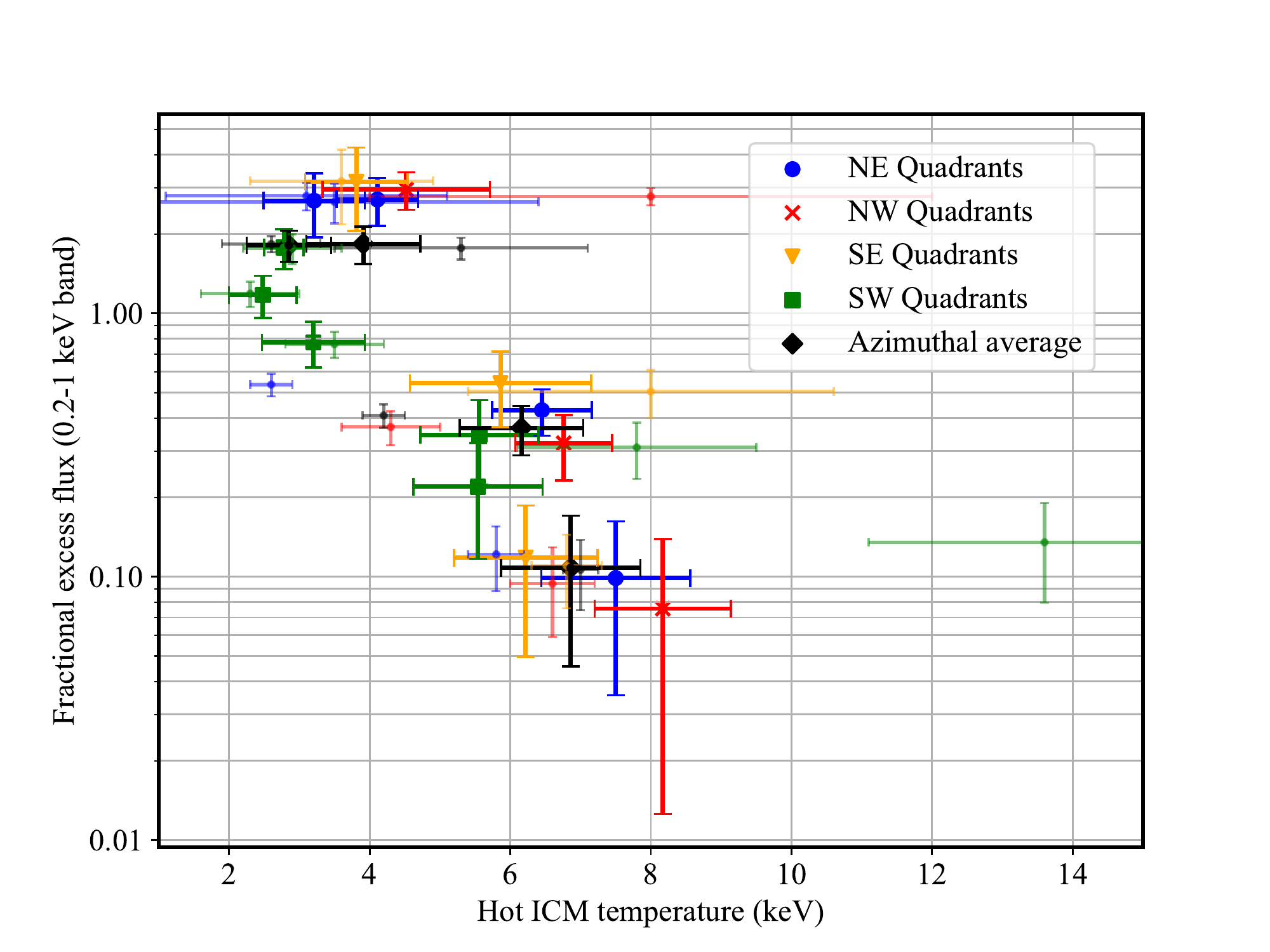}
  \caption{Distribution of fractional soft excess fluxes as a function
  of plasma temperature. For comparison, lighter--color datapoints represent the results using the 
  original \protect\cite{bonamente2003} temperatures.}
  \label{fig:etakT}
\end{figure*}

\begin{figure*}
  \includegraphics[width=5.5in]{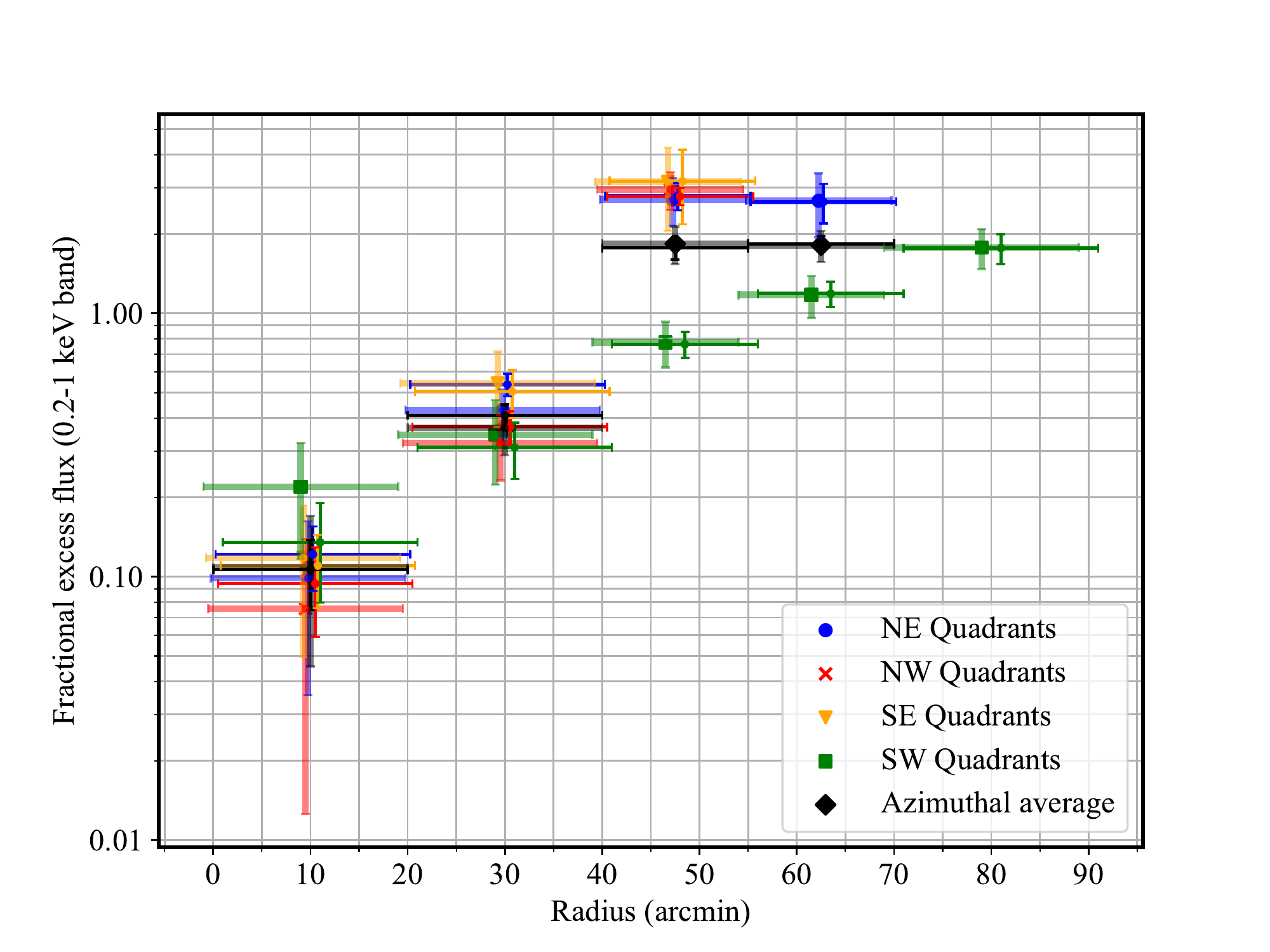}
  \caption{Distribution of fractional soft excess fluxes as a function
  of radial distance from the center of the Coma cluster. Lighter--color datapoints represent the results using the
  original \protect\cite{bonamente2003} temperatures. Datapoints at the same radius where slightly shifted for clarity.}
  \label{fig:etaR}
\end{figure*}

The result of re--scaling the predicted soft X--ray fluxes according to Sect.~\ref{subsec:rescaling}
are presented in Table~\ref{tab:rescale} and Figure~\ref{fig:etakT}, where the fractional
soft excess flux is defined as 
\begin{equation}
  \eta = \dfrac{F_{24}-P_{24}}{P_{24}}
  \label{eq:eta}
\end{equation}
in the same manner as in \cite{bonamente2003}. Figure~\ref{fig:etaR} also shows the distribution of the
soft excess fluxes as a function of radial distance from the center of the Coma cluster, 
{using the same $\eta$ values
as in Figure~\ref{fig:etakT}}.

\begin{table*}
  \caption{Soft excess fluxes using the new \xmm/\planck\ temperatures. $P_{24}$ are the original predicted
  fluxes from \protect\cite{bonamente2003} (the small errors are ignored), $\Delta P_{24}$ is the change based on the temperature change $\Delta kT$ between the
  new \xmm/\planck\ temperature and the old \rosat\ temperatures, and $\eta$
  are the fractional soft excess fluxes.}
  \label{tab:rescale}
  \begin{tabular}{llccccc}
    \hline
    \hline
    Quadrant & Radius & $\dfrac{\Delta kT}{kT}$ & $P_{24}$ &  $\Delta P_{24}$ & $\eta$ & $\eta_{\new}$ \\
       & (arcmin)   &  ---                   & (c~s$^{-1}$) & (c~s$^{-1}$) & --- & --- \\
    \hline
    NE &  $10.0\pm10.0$ & 0.29 & 1.810 & $0.038 \pm 0.044$ & $0.122 \pm 0.034$  & $0.099 \pm 0.063$\\
NE &  $30.0\pm10.0$ & 1.48 & 0.345 & $0.026 \pm 0.009$ & $0.536 \pm 0.053$  & $0.429 \pm 0.085$\\
NE &  $47.5\pm7.5$ & 0.33 & 0.095 & $0.002 \pm 0.010$ & $2.789 \pm 0.332$  & $2.705 \pm 0.559$\\
NE &  $62.5\pm7.5$ & -0.08 & 0.137 & $-0.001 \pm 0.018$ & $2.650 \pm 0.454$  & $2.675 \pm 0.732$\\
NW &  $10.0\pm10.0$ & 0.24 & 1.700 & $0.029 \pm 0.039$ & $0.094 \pm 0.035$  & $0.076 \pm 0.063$\\
NW &  $30.0\pm10.0$ & 0.57 & 0.500 & $0.018 \pm 0.015$ & $0.370 \pm 0.054$  & $0.321 \pm 0.089$\\
NW &  $47.5\pm7.5$ & -0.44 & 0.180 & $-0.008 \pm 0.015$ & $2.778 \pm 0.207$  & $2.954 \pm 0.482$\\
SE &  $10.0\pm10.0$ & -0.09 & 2.180 & $-0.015 \pm 0.058$ & $0.110 \pm 0.034$  & $0.118 \pm 0.068$\\
SE &  $30.0\pm10.0$ & -0.27 & 0.420 & $-0.010 \pm 0.024$ & $0.505 \pm 0.104$  & $0.543 \pm 0.173$\\
SE &  $47.5\pm7.5$ & 0.06 & 0.091 & $0.000 \pm 0.006$ & $3.176 \pm 0.997$  & $3.157 \pm 1.103$\\
SW &  $10.0\pm10.0$ & -0.59 & 2.000 & $-0.139 \pm 0.069$ & $0.135 \pm 0.055$  & $0.220 \pm 0.103$\\
SW &  $30.0\pm10.0$ & -0.29 & 0.714 & $-0.019 \pm 0.028$ & $0.310 \pm 0.075$  & $0.345 \pm 0.122$\\
SW &  $47.5\pm7.5$ & -0.09 & 0.380 & $-0.003 \pm 0.017$ & $0.763 \pm 0.086$  & $0.776 \pm 0.154$\\
SW &  $62.5\pm7.5$ & 0.08 & 0.160 & $0.001 \pm 0.009$ & $1.188 \pm 0.131$  & $1.174 \pm 0.214$\\
SW &  $80.0\pm10.0$ & -0.04 & 0.121 & $-0.000 \pm 0.005$ & $1.769 \pm 0.231$  & $1.778 \pm 0.307$\\

    \hline
     & $10.0\pm10.0$ & -0.02 & 8.000 & $-0.013 \pm 0.178$ & $0.106 \pm 0.032$ & $0.108 \pm 0.062$\\
 & $30.0\pm10.0$ & 0.47 & 1.880 & $0.058 \pm 0.046$ & $0.410 \pm 0.042$ & $0.367 \pm 0.078$\\
 & $47.5\pm7.5$ & -0.26 & 0.650 & $-0.016 \pm 0.038$ & $1.769 \pm 0.168$ & $1.837 \pm 0.299$\\
 & $62.5\pm7.5$ & 0.10 & 0.300 & $0.002 \pm 0.015$ & $1.833 \pm 0.129$ & $1.813 \pm 0.245$\\

    \hline
    \hline
  \end{tabular}
\end{table*}

Uncertainties in the new R24 band predictions and in the fractional soft excess fluxes are obtained
through standard error propagation formulas \citep[e.g., see Chapter 5 of][]{bonamente2021book}.
According to \eqref{eqc}, the standard error in $z=P_{24,\new}$ is given by
\[ \dfrac{\sigma^2_z}{z^2} = \left(\dfrac{1}{2}+b\right)^2 \left(\dfrac{\sigma^2_{x}}{x^2} + \dfrac{\sigma^2_{y}}{y^2} \right)^2\]
where $x=kT$ and $y=kT_{\new}$ are the \rosat\ and \xmm/\planck\ temperatures of Table~\ref{tab:tprofComp}, respectively.
Likewise, the error in the fractional soft excess fluxes according to \eqref{eq:eta} is given by
\[ \sigma^2_{\eta} = \left(\dfrac{F}{P} \right)^2\left( \dfrac{\sigma^2_F}{F^2} + \dfrac{\sigma^2_P}{P^2}\right)^2 \]
where $F$ is the measured \rosat\ R24 band flux, which remained constant throughout the analysis, 
and $P$ is the predicted flux in that band, indicated as respectively $P_{24}$ for the \rosat\ temperatures, and
$P_{24,\new}$ for the \xmm/\planck\ temperatures.
The results of Table~\ref{tab:rescale} illustrate that the weak dependence of the hot ICM--predicted soft X-ray fluxes $P_{24}$ on temperature,
according to \eqref{eqc}, leads to only small changes in soft excess fluxes between the cases of \rosat\ or \xmm/\planck\ temperatures. 

{The small effect of the hot ICM temperature on soft excess fluxes also indicates that
any multi--phase structure along the sight line (i.e., in a given annular region) is unlikely
to have a significant effect on our results. We therefore regard the best--fit temperature
in an given region as the average temperature along the sightline, and do not include systematic errors
associated with possible multi--temperature gas.}


\begin{figure}
  \includegraphics[width=3.5in]{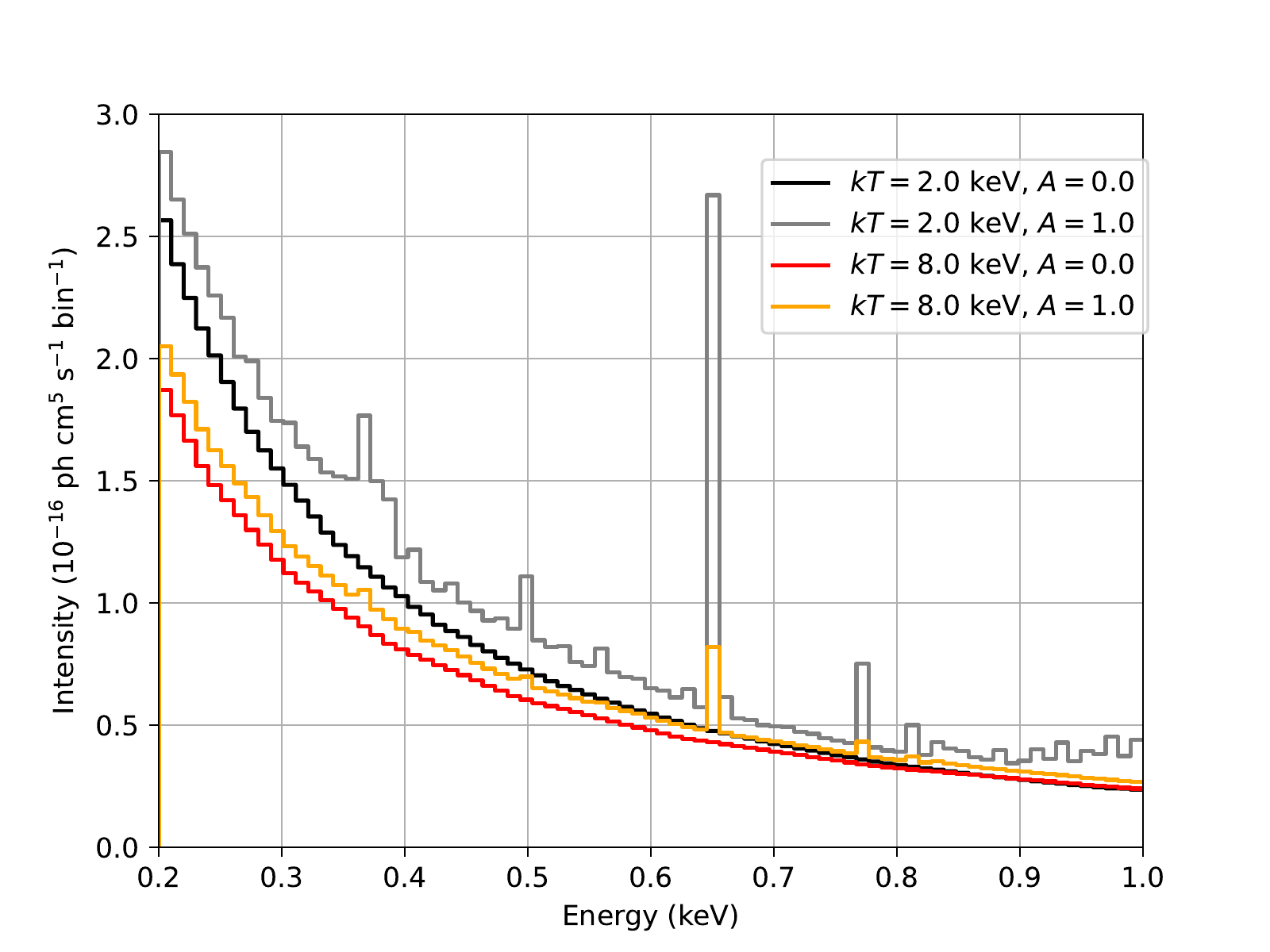}
  \caption{Spectral intensity of an \texttt{APEC} optically--thin thermal spectrum in collisional ionization equilibrium
  at two characteristic hot ICM temperatures. Continuum emission is the dominant mechanisms at these temperatures, with line emission becoming more significant at lower temperatures.}
  \label{fig:APEC}
\end{figure}
Both the original \rosat\ analysis and the newer \xmm/\planck\ results assumed a significantly sub--solar abundance of 
$Z>2$ elements. In particular, at large radii the \rosat\ data were analyzed with a \texttt{MEKAL} model with $A=0.2$ Solar
abundances, and the \xmm/\planck\ data with a \texttt{APEC} thermal model with $A=0.3$ Solar abundances.
To address the effect of uncertainties in our knowledge of the chemical abudance of the hot ICM in Coma, we calculated the spectral
intensity of an optically--thin thermal spectrum in collisional equilibrium, for temperatures in the range $kT=2-8$~KeV, which
are typical of the Coma cluster. Figure~\ref{fig:APEC} shows the spectral intensity of the optically--thin \texttt{APEC} emission model \citep{smith2001},
which is a higher--resolution replacement for the \texttt{MEKAL} model.~\footnote{The \texttt{APEC} model is accessiblee
via the \texttt{ATOMDB} project, which also provides the \texttt{PyAtomDB} software at \url{https://atomdb.readthedocs.io/en/master/index.html}.}
At a temperature of $kT=2$~keV, the total spectral intensity of an optically--thin \texttt{APEC} plasma with
Solar abundances of $Z>2$ elements  \citep[e.g., according to][]{anders1989} is $7.56\times 10^{-15}$~photons~cm$^{-5}$~s$^{-1}$,
with 24.7\% of the total intensity provided by line emission. At a temperature of $kT=8$~keV, lines provide only
11.2\% of the intensity, consistent with an increased dominance of continuum emission at higher temperatures, where
most of the ICM is fully ionized. These calculations clearly indicate that small changes in the metal abundance (e.g., of order $\Delta A=0.1$),
only contribute to a small ($\sim 1-2$~\%) change in the plasma emissivity, and in the associated prediction of the R24 band fluxes
from an hot ICM at these temperatures. To account for our uncertainty in the chemical abundances in the Coma hot plasma, we introduce
a systematic error of $\pm 2\%$ in $P_{24,\new}$ to the calculation of the errors in the fractional soft excess fluxes.
These errors are included in the $\eta_{\new}$ column of Table~\ref{tab:rescale}.

\section{Confirmation of strong soft excess emission in Coma and its astrophysical implications}
\label{sec:results}

The results of Table~\ref{tab:rescale} show that the soft excess emission is detected with high statistical
significance ($\geq 3 \sigma$) in all regions at $r \geq 20$~arcmin, as also shown in Figure~\ref{fig:etaR}.
The excess becomes stronger, relative to the emission of the hot ICM, at large radial distances from the center
of the Coma cluster. Use of the more accurate \xmm/\planck\ temperatures, and an allowance for systematic errors
associated with the metal abundances, therefore confirms the results of \cite{lieu1996b} and of \cite{bonamente2003,bonamente2009}
of strong soft excess emission from Coma.

Although the spectral resolution of the soft excess emission in the 0.2-1~keV band is limited, the
previous \rosat\
analysis suggested that a thermal origin for the soft excess, as modelled via a single--temperature additional thermal component,
is to be preferred to
a non--thermal origin, as modelled with a power--law component. It is therefore instructive to follow the thermal
interpretation for the soft excess signal and estimate the inferred warm gas masses. 
The \rosat\ spectral modelling of the soft excess indicated a warm gas temperature of $\log T(\mathrm{K}) \simeq 6-6.5$,
with a metal abundance $A\leq 0.3$, in all regions. These conditions are also consistent with a significant
fraction of the warm--hot intergalactic medium (WHIM) that is predicted in large--scale filaments that converge
towards clusters \citep[e.g.][]{wijers2019,tuominen2021}. From these observations, it is not possible
to determine whether the putative warm gas is \emph{within} the cluster, or whether it is seen 
\emph{in projection}
onto the cluster. We therefore follow both scenarios to provide an estimate of the mass that could be responsible for this signal
in Sects.~\ref{sec:icm} and \ref{sec:whim},
and we draw our conclusion from these estimates in Sect.~\ref{sec:icmvswhim}.

\subsection{Mass estimates for warm \emph{intra--cluster} gas}
\label{sec:icm}
{The soft excess emission can in principle originate from warm, sub--virial intra--cluster gas.
In this scenario, the warm gas would likely be clumpy 
\citep[i.e., with a volume filling factor $f<1$, as discussed in Sect.~3 of ][]{bonamente2001}, in order to maintain local pressure equilibrium with the hotter ICM. For the calculations
of this section we assume no clumping ($f=1$), and comment on how 
the results would be modified in the presence of clumping at the end of the section.}

For each quadrant, the soft excess flux $S_{24} = F_{24}-P_{24}$ is first re--scaled according to the new \xmm/\planck\
temperatures, according to the method of Sect.~\ref{sec:rescaling}. This flux is proportional to the emission integral $I_w$
of the warm gas, with the characteristics reported in Table~3 of \cite{bonamente2003} (viz., the emission integral, temperature and
abundance of the warm gas model). 
Accordingly, the measured emission integrals of the warm gas are rescaled in proportion
to their change in $S_{24}$ for the new temperatures of the hot gas.

Moreover, we examined the possibility that the warm gas temperatures may vary from the best--fit values of the \rosat\ analysis,
given the limited spectral resolution of {PSPC}.
To this end, we used the \texttt{ATOMDB} data for the \texttt{APEC} thermal model to determine the changes in the warm gas
emissivity as a function of both the temperature and abundance. Unlike the case of the hot gas at $\log T(\mathrm{K})\geq 7$ (see Fig.~\ref{fig:APEC}), 
the emissivity of warm gas at $\log T(\mathrm{K})=6-7$ depends strongly on metal abundances, since there is a much larger
contribution from line emission than at higher temperatures, where most of the ions are fully ionized, in collisional
ionization equilibrium \citep[e.g.][]{mazzotta1998}. The warm gas emission integrals were therefore 
also rescaled in proportion to the change
in the emissivity of the warm plasma according to \eqref{eq:I}, 
for a few representative abundances and temperatures of the warm gas ($\log T(K)=6,7$
and $A$=0.1,1).

In each spectral region, the ratio of warm--to--hot gas mass is given by
\begin{equation}
  \dfrac{M_{w,i}}{M_{h,i}} = \dfrac{\overline{n_{w,i}}\; V_i}{\overline{n_{h,i}}\; V_i} = \left(\dfrac{I_{w,i}}{I_{h,i}}\right)^{\nicefrac{1}{2}}
  \label{eq:mratio}
\end{equation}
where $V_i$ is the volume of the representative region, $\overline{n}$ indicates average density, $I_h$ is the emission integral of the hot plasma,
and $I_w$ that of the warm plasma, the latter being rescaled for different choices of the warm gas temperature and abundance
according to \eqref{eq:I} as described above.
We used this relationship  to calculate the
warm gas mass in each annular region covered by the \cite{mirakhor2020} analysis (see their Figure~7), using the
ratio of emission integrals of the warm--to--hot gas from the \rosat\ data. Since the \xmm/\planck\ masses were reported 
on a finer angular resolution than the \rosat\ soft excess fluxes, the ratios according to \eqref{eq:mratio}
were constant for a few of the  \xmm/\planck\ differential mass bins. 
Uncertainties in the warm gas masses are obtained by a standard error propagation based on \eqref{eq:mratio},
where for simplicity the smaller uncertanties of the emission integrals of the hot phase were ignored.

\begin{figure*}
  \centering
  \includegraphics[width=5in]{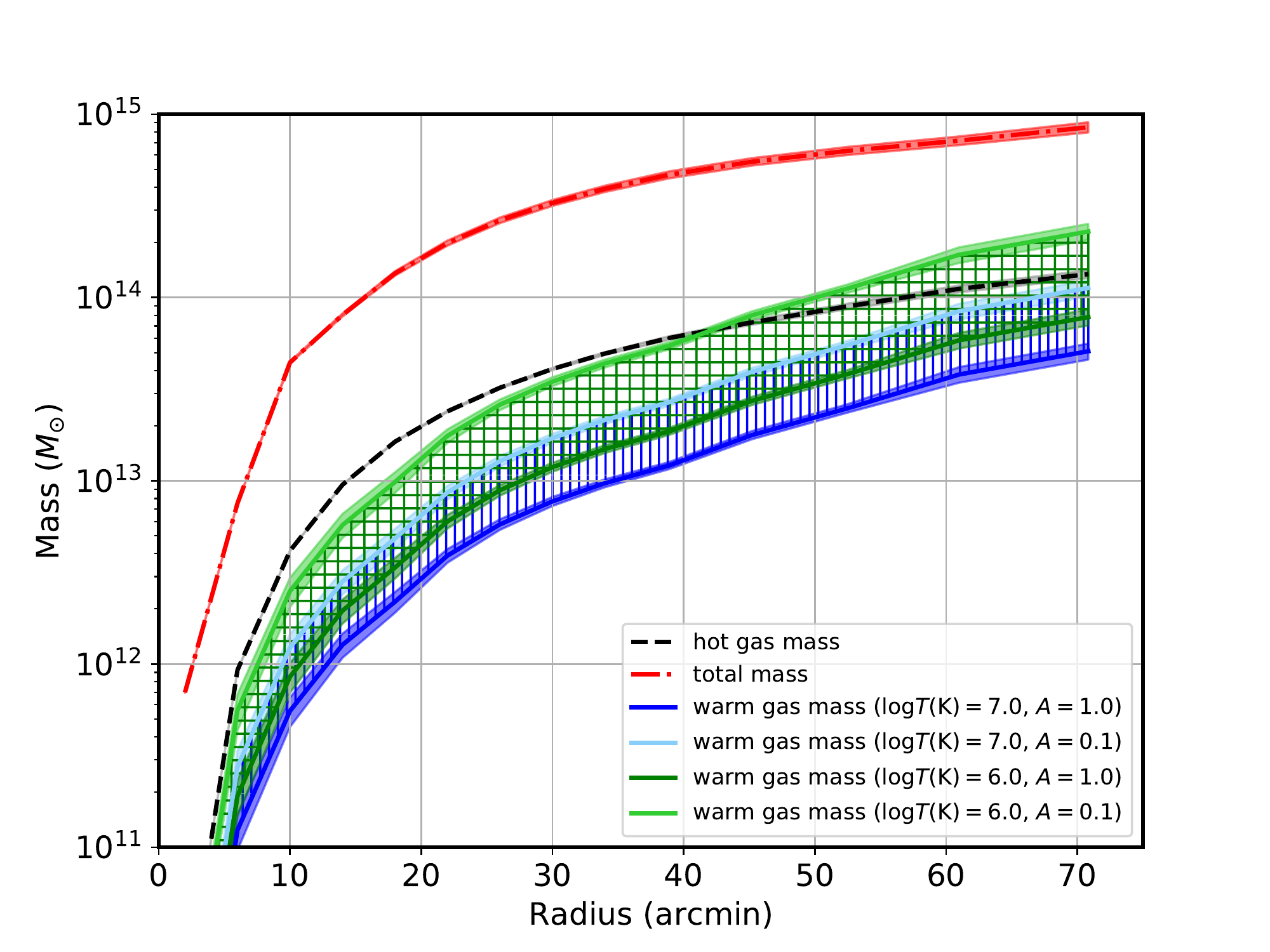}
  \caption{Estimated masses in the warm phase of the intracluster medium, according to the 
  soft excess flux in the Coma cluster, for different choices of the temperature and abundance
  of the warm gas. Confidence bands are at the $1\,\sigma$ level. Hatched areas indicate the range of warm gas masses
  for two representative temperatures, when the metal abundances vary from 10\% to 100\% Solar. }
  \label{fig:masses}
\end{figure*}

The results of the estimates are shown in Figure~\ref{fig:masses}, 
where in blue are two estimates obtained for $\log T(K)=7$ and metal abundances of respectively 10\% and 100\% Solar, 
and in green two similar estimates for a lower temperature of $\log T(K)=6$.
As expected, the estimated  gas masses are strong function of both the temperature and abundaces assumed for the warm gas.
For a metal abundance of 0.1 Solar, the emissivity in the 0.2-1~keV band is 
$1.1\times 10^{-24}$~erg~cm$^{-3}$~s$^{-1}$ at $\log T(K)=6$, 
increasing approximately four times to to $4.7\times 10^{-24}$~erg~cm$^{-3}$~s$^{-1}$ at a temperature of $\log T(K)=7$. 
This increase in emissivity corresponds to a nearly two--fold  \emph{decrease} in estimated gas masses, consistent with
the comparison between the light blue and light green mass profiles in Fig.~\ref{fig:masses}.
Similar considerations apply to the comparison of warm gas mass estimates for the same temperature, but for varying metal abundances.

The main result from the intra--cluster warm gas interpretation of Fig.~\ref{fig:masses} 
is that the warm gas mass inferred from the soft excess fluxes
are comparable to that of the hot gas, and they are strongly dependent on the uncertain physical
conditions of the warm gas. It is also clear that, consistent with the increasing fraction of soft excess fluxes towards
large radii, the putative warm gas becomes dominat towards the cluster's virial radius.
{The intra--cluster warm gas masses would be reduced according to $M \propto \sqrt{f}$ in the presence of clumping ($f<1$),
since the emitting gas would be denser, and the X--ray 
emission is proportional to the square of the density \citep[e.g.][]{bonamente2001}.}

\subsection{Warm gas mass estimates for a WHIM origin of the soft excess}
\label{sec:whim}
\begin{figure*}
  \centering
  \includegraphics[width=5in]{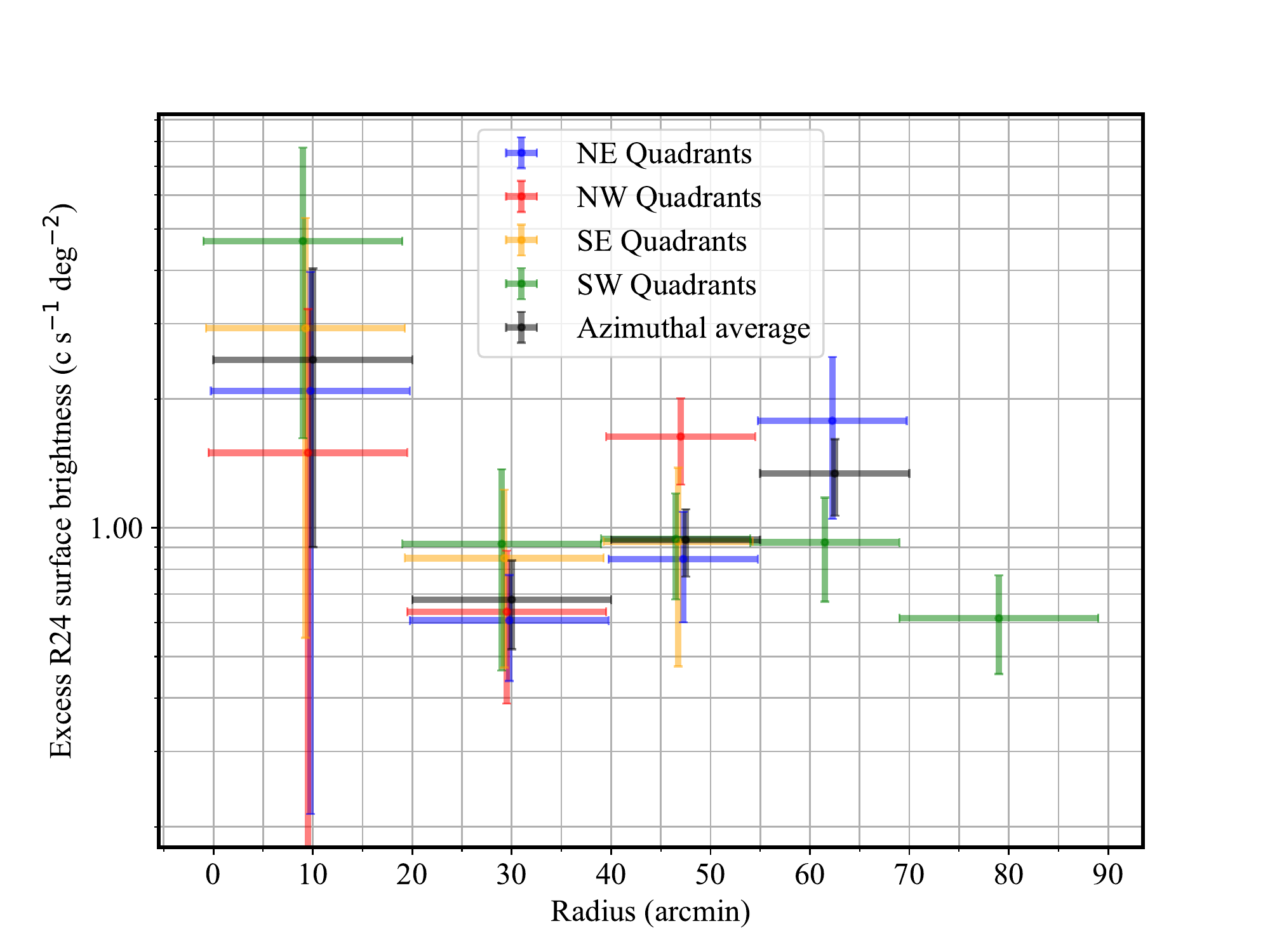}
  \caption{Surface brightness of the soft excess in each of the \rosat\ regions. The soft excess fluxes
  $F_{24}-P_{24}$ were divided by the area of each region (quadrant or annulus).}
  \label{fig:whim}
\end{figure*}

To investigate whether the soft excess emission could originate from WHIM filaments projected towards Coma,
Figure~\ref{fig:whim} provides the radial profile of the surface brightness of the soft excess.
For this profile, the excess flux $F_{24}-P_{24}$, rescaled for the \xmm/\planck\ temperatures,
was divided by the area of the extraction region, accounting for excluded regions (e.g., point sources
excised from the field of view, and partial coverage of the annuli or quadrants).
The results show that  the surface brightness is rather \emph{constant} with radius, providing general
consistency with a simple model where approximately uniform--density WHIM filaments
are seen in projection towards the cluster. While WHIM filaments are expected to have gradients in their
WHIM density \citep[see, e.g.][]{tuominen2021}, these gradients are much milder than those of the hot ICM gas density,
which typically vary from the center to the virial radius of a cluster by more than two orders of magnitude.

If the soft excess originates from WHIM filaments projected onto Coma, it is possible
to estimate the length and mass associated with the filaments, as a function of
the filament density. The estimate is based upon the relationship that applies to each spectral extraction region
\def\nwhim{n_{\mathrm{WHIM}}}
\begin{equation}
  I_w = \int n^2 dV \simeq  \overline{n}_{\mathrm WHIM}^2 \cdot (L_{\mathrm WHIM} \times A),
  \label{eq:whim}
\end{equation}
where $I_w$ is the measured emission integral of the warm gas, $A$ is the area, $L_{\mathrm WHIM}$ 
is the length of the filament, and $\nwhim$\ is a characteristic density for the
WHIM that needs to be assumed a priori. Gas in a dense WHIM filament is expected to 
feature baryon overdensities
\[\delta_{\rho}= \dfrac{\rho_{b}-\overline{\rho}_b}{\overline{\rho}_b} \sim 30-300,\]
where 
\[ \overline{\rho}_b=4.2 \times 10^{-31}~\text{g cm}^{-3} = 2.6 \times 10^{-7} \text{ H atoms}~\mathrm{cm}^{-3}\]
is the mean baryon density of the
universe, for the standard $\Lambda$CDM \planck\ cosmology \citep{tuominen2021,holt2022,planckCosmology2014,planck2016-cosmology}.

According to  \eqref{eq:whim}, all \rosat\ regions can be interpreted with a WHIM origin of the soft
excess, where filaments have characteristic lengths in the range
\[ L_{\mathrm WHIM} = (4.5 - 33.3) \times \left(\dfrac{n_{\mathrm WHIM}}{7.8 \times 10^{-5}~\mathrm{cm}^{-3}} \right)^{-2} \text{ Mpc}
\]
where $n_{\mathrm WHIM}=7.8 \times 10^{-5}~\mathrm{cm}^{-3}$ corresponds to $\delta_{b}=300$, for an average length of
\[  L_{\mathrm WHIM} = 15.5\pm9.1  \times \left(\dfrac{n_{\mathrm WHIM}}{7.8 \times 10^{-5}~\mathrm{cm}^{-3}} \right)^{-2} \text{ Mpc}.
\]
These lengths are generally consistent with WHIM filaments seen in \texttt{EAGLE} simulations 
\citep[e.g.][]{wijers2019, tuominen2021,holt2022}, and the region--to--region scatter can be attributed to
mild gradients in the WHIM density and to the geometry of the filaments. 
For these estimates, we assumed the original emission integrals
for the warm phase as reported in \cite{bonamente2003}, without any rescaling by temperaure and abundance
as was done in Sect.~\ref{sec:icm} for the intra--cluster interpretation of the soft excess. 
Any such rescaling would result in changes 
to these estimates by a factor of few, according to \eqref{eq:I} and \eqref{eq:whim}, and thuse leave all
considerations provided in this section largely unchanged.
Notice that, if the filaments where to have a significantly lower density, e.g., $\delta_{\rho}=30$, 
the filament length estimates 
according to \eqref{eq:whim} would be significantly larger, i.e., by a factor $\times 100$. Such filament
lengths would be at odds with our current understanding of the large--scale filamentary structures.
Our data therefore suggest that WHIM filaments towards Coma are indeed on the high--density end of the range
expected based on the \texttt{EAGLE} numerical simulations. 

Following this WHIM interpretation with $\delta_{\rho}=300$ filaments, the warm gas mass in the \rosat\
regions is estimated as
\[ M_{\mathrm WHIM} = 2.2\pm0.2\times 10^{14} \left(\dfrac{n_{\mathrm WHIM}}{7.8 \times 10^{-5}~\mathrm{cm}^{-3}} \right)^{-1} M_{\odot},\]
where the inverse dependence on the assumed WHIM density
derives from the inverse proportionality between filaments lengths and the square of the
density, according to \eqref{eq:whim}, and the error is the statistical uncertainty associated with the
measured emission integrals of the warm gas. This mass estimate does not include any region not explicitly covered 
by the \rosat\ observations, and therefore is to be considered as a strict lower limit, especially given that the
coverage
at large radii is very sparse. In this paper we do not attempt to correct for the partial coverage of the \rosat\ observations,
which would lead to an increase of these mass estimates possibly by a factor of few times.

This mass estimate is generally consistent with
the \texttt{EAGLE} simulation results of \cite{tuominen2021} and \cite{holt2022}, 
where in a $100^3$~Mpc$^{3}$ volume there is a total {\bf mass}  of $2.2 \times 10^{15}$~$M_{\odot}$
in several WHIM filaments in the temperature range $\log T(K) = 5-7$, as identified by the
\emph{Bisous} method \citep{stoica2007}. Our observations of the soft excess emission in the 
Coma cluster are therefore consistent with a WHIM origin, where WHIM filaments with
density, temperature and size of the kind seen in the \texttt{EAGLE} simulations, are
responsible for the excess of soft X--ray radiation detected by \rosat.

\subsection{Intra--cluster vs. WHIM origin of the excess}
\label{sec:icmvswhim}

The intra--cluster origin of the soft excess was investigated in the hydrodynamical
simulations of \cite{cheng2005}. Their results indicated that an excess of soft X--ray
radiation can  originate from low--entropy, high--density cooler gas within the virial
radius of the cluster. Within this scenario, the gas responsible for the soft excess 
is inhomogenous or clumpy {(as also discussed in Sect.~\ref{sec:icm})}, as opposed 
to the more diffuse and homogeneous nature for the hotter, lower--entropy
hot ICM. Such clumpiness implies that mass estimates based on a diffuse gas (as those
in Sect.~\ref{sec:icm}) would be overestimated by a factor equal to the degree of 
clumpiness of the gas, defined as
\[
  C = \dfrac{\overline{n^2}}{\overline{n}^2} = \dfrac{1}{f} \geq 1,
\]
{where the volume of the emitting gas is $V_g = f \times V$, with $V$ the 
physical volume of
the spectral region and $f \leq 1$ the volume filling factor.}
\cite{cheng2005} also found that the amount of soft excess of Coma cannot be
readily interpreted as intra--cluster clumpy warm gas, which was seen 
in lower amounts than those required to explain the Coma excess. The excess
emission seen in other clusters, e.g., in the sample studied in \cite{bonamente2002},
was at a lower level, and one that is more consistent with the  \cite{cheng2005} simulations.
{Another problem with the intra--cluster warm gas scenario is that such gas would
have a short cooling time, raising the issue of its maintenance over a cosmological
time \citep[e.g.][]{bonamente2001}.}

Large--scale hydrodynamical simulations from \texttt{EAGLE}, on the other hand, feature
WHIM filaments that can directly explain the soft excess emission in the Coma cluster.
In this scenario, filaments with $\sim 10-20$~Mpc length, a temperature of $\log T(K)\sim 6$ and an average  baryon overdensity 
$\delta_{\rho} \sim 300$, provide a natural interpretation for the excess of soft X--ray radiation, which is
seen in projection against the cluster, {as described above in Sect.~\ref{sec:whim}}. 
A similar picture also emerges from the \texttt{IllustrisTNG}
simulations \citep[see, e.g.][]{martizzi2019,gallarraga2021}, where warm WHIM gas  is also preferentially
found at large projected distances from the cluster core \citep{gouin2022}. 

Additional observational evidence of a WHIM origin for the soft excess in Coma is provided by the
spatial distribution of the hot gas and by the measurements of the gas mass fraction and entropy profiles by \cite{mirakhor2020}. 
In fact, along the south--western quadrant where
Coma connects with Abell~1367, there is a well--known enhancement of X--ray brightness
accompanied by a gas mass fraction that is significantly above the cosmic mean,
also featuring a decrease in entropy (see bottom--right panels of Figs.~11 and 13 of \citealt{mirakhor2020}).
As shown in Figs.~\ref{fig:etakT}--\ref{fig:etaR} and in Fig.~\ref{fig:whim}, this quadrant also has significant
soft excess emission that extends  beyond Coma's virial radius. A natural interpretation
of the higher--than--cosmic  gas mass fraction from the \xmm/\planck\ 
data is that there is an especially hot filament between
Coma and Abell~1367, with significant emission both in the main and in the soft X--ray bands. 
Such filament, when
projected against the more spherically--symmetrical distribution of the 
cluster hot gas and its underlying dark matter potential, could explain the overall higher gas mass fraction in that quadrant.
In other azimuthal directions where the gas mass fraction reaches the cosmic value
near the virial radius (see other panels of Fig.~13 of \citealt{mirakhor2020}), the soft X--ray excess is plausibly provided by WHIM filaments at lower
temperatures that emit primarily in the soft X--ray band, and thus providing little or no
contribution to the gas mass fraction. 
The presence of these filaments is also consistent with the predictions of numerical simulations
by \cite{zinger2016}, where low--entropy WHIM filaments are seen to penetrate deep into the cluster interior.
If the soft excess
had originated from an intra--cluster gas instead, its large mass implications (see Fig.~\ref{fig:masses}) 
would further increase the gas mass fraction in all quadrants to significantly higher--than--cosmic values,
especially in the south--western quadrant, thus exacerbating the dicrepancy between the gas mass fraction and the cosmological
ratio of baryons to total matter. 

The concurrence of 
(a)  quantitative agreement between the
estimated temperature, mass and filament lengths inferred from the observed soft excess with the predictions from numerical simulations
of the WHIM, and (b) difficulties in the interpretation of the soft exces as intra--cluster gas, especially in terms of the gas mass fraction,
leads to the conclusion that the most likely origin for the cluster soft excess is a WHIM origin.

\section{Discussion and conclusions}
\label{sec:conclusions}

The availability of high--resolution X--ray measurements of Coma's temperature profile
by \xmm\ and \planck\ \citep{mirakhor2020} have provided confirmation of the presence of strong
soft excess emission associated with the cluster.
The soft excess was detected in  \rosat\ data
\citep{bonamente2003}, and it is consistent with a thermal origin from warm gas at sub--virial
temperatures, $\log T(K) \leq 7$. A similar excess of soft X--ray radiation  above the contribution
from the hot ICM is present in several other clusters \citep[e.g.][]{bonamente2002}, but the
size, distance and brightness of Coma provide a unique opportunity for the investigation of this
phenomenon. In particular, we have investigated two possible origins
for the radiation: an intra--cluster origin, where the warm gas co--exists with the hot gas,
and a WHIM origin, where the gas is located in filaments that converge towards the cluster.
The agreement between expected properties of the WHIM and those inferred from the soft excess, and
the difficulties in the interpretation of the excess as warm intra--cluster gas, lead to the conclusion that
the most likely interpretation for the soft excess is emission from WHIM filaments that converge towards the
massive Coma cluster.

There have been a wealth of observational efforts towards the 
direct detection of WHIM filaments.
The more commonly used methods include the stacking of SZE signal \citep[e.g.][]{degraaf2019,tanimura2019} and
the direct detection of its X--ray emission
\citep[e.g.][]{kull1999,eckert2015,scharf2000,werner2008,zappacosta2005}, in addition to the indirect method
to identify WHIM filaments via X--ray \citep[e.g.][]{bonamente2016,nicastro2018}
or FUV absorption lines \citep[e.g.][]{danforth2016,tilton2012}. 
The soft excess phenomenon, demonstrated with high statistical significance in this paper for the Coma cluster,
provides a unique way to overcome some of the limitations of the other 
direct methods of detection of the WHIM. In fact, WHIM filaments are expected to be denser 
when seen -- in projection --  within a cluster's virial radius, and therefore with stronger X--ray emission
than when seen beyond that radius \citep[e.g., as in the case of ][]{eckert2015}, or for filaments that are not
associated with massive clusters \citep[e.g., in ][]{degraaf2019}.

It remains an open question whether the WHIM can solve the \emph{missing baryons} problem, i.e., the
observation that a significant portion (approximately $\nicefrac{1}{2}$) of the low--redshift baryons 
have not been identified yet \citep[e.g.][]{danforth2016}.
Both numerical simulations and observations are converging on a consesus that WHIM baryons in filaments are
sufficient to bridge the missing baryons gap. The ultimate confirmation of the resolution of this problem
can only be provided observationally, and the current results are certainly encouraging, but based only on
a small sampling of the local universe, as in the cases of \cite{nicastro2018} and \cite{eckert2015}.
The cluster soft excess phenomenon offers a unique opportunity to observe a large sample of nearby clusters
with current and future missions featuring stable and well--calibrated soft X--ray response, such as \texttt{eRosita}, in order to
conclusively solve this problem.

\section*{Acknowledgements}
MSM and SAW acknowledge support from the NASA XMM-Newton grant 19-XMMNC18-0030.

\section*{Data availability}
The \rosat\ tabular data underlying this article are provided in the 
\cite{bonamente2003} paper, available from the publisher at \url{https://iopscience.iop.org/article/10.1086/346220}.
The \xmm\ and \planck tabular data underlying this article are accessible from the
\cite{mirakhor2020} paper, available from the publisher at
\url{https://academic.oup.com/mnras/article/497/3/3204/5876891}.
The \rosat\ data are also available from the \texttt{HEASARC} archive at \url{https://academic.oup.com/mnras/article/497/3/3204/5876891}.
The \xmm\ data are available from the \texttt{XSA} archive available at \url{https://www.cosmos.esa.int/web/xmm-newton/xsa#access}.
The \planck\ data underlying this paper can be accessed from the archive available at \url{https://www.cosmos.esa.int/web/planck}.
Processed data from this paper are available upon request to the authors.






\bibliographystyle{mn2e}
\bibliography{/home/max/proposals/max}


\bsp	
\label{lastpage}
\end{document}